# Threat Mitigation: The Asteroid Tugboat
(White paper 041)

Russell Schweickart, Clark Chapman, Dan Durda, Piet Hut[1]

Summary: The Asteroid Tugboat (AT) is a fully controlled asteroid deflection concept using a robotic spacecraft powered by a high efficiency, electric propulsion system (ion or plasma) which docks with and attaches to the asteroid, conducts preliminary operations, and then thrusts continuously parallel to the asteroid velocity vector until the desired velocity change is achieved. Based on early warning, provided by ground tracking and orbit prediction, it would be deployed a decade or more prior to a potential impact. On completion of the initial rendezvous with the near-Earth object (NEO) the AT would first reduce the uncertainty in the orbit of the asteroid via Earth tracking of its radio transponder while it is station keeping with the asteroid. If on analysis of tracking data a deflection is required the AT would execute a reconnaissance phase collecting and processing information about the physical characteristics of the asteroid to support subsequent operations. The AT would then dock at the appropriate pole (i.e. on the spin axis), attach to the asteroid surface, and initiate a NEO reorientation maneuver. Following completion of the NEO reorientation the AT would initiate the deflection phase by thrusting continuously parallel to the asteroid velocity vector until the resultant target orbit is achieved. The orbit of the asteroid is continuously monitored throughout the deflection process and the end state is known in real time. The performance capability of the AT, since it uses the high performance electric propulsion system of the spacecraft for both NEO rendezvous and the subsequent deflection, depends on the future development of this technology. If one assumes a nuclear-electric propulsion (NEP) system similar to that formerly under development in the recently canceled Prometheus Program, the AT would be capable of deflecting threatening NEOs up to 800 meters in diameter or more.

## I Introduction

The first public presentation of the Asteroid Tugboat (AT) concept was its publication in the November, 2003 issue of Scientific American magazine[2], authored by Russell L. Schweickart, Edward T. Lu, Piet Hut, and Clark R. Chapman. The AT is a conceptually straightforward design which deflects a threatening near-Earth asteroid (NEO) by docking with the asteroid a decade or more prior to the nominal impact and pushing it gently in the proper direction to affect an orbital modification sufficient to cause the asteroid to miss the Earth.

The deflection process is controlled throughout with precise radio transponder determination of the asteroid orbit on arrival and continuous tracking of the asteroid and docked spacecraft during and after the deflection maneuver. A specific target orbit for the asteroid is determined prior to execution of the deflection phase and the maneuver is controlled during deflection to achieve this targeted end point.

In the general case an asteroid deflection mission will be called for when an asteroid has been discovered, tracked and determined to have a significant probability of impact with the Earth. The specific timing for the deployment of a deflection mission will depend on many factors, among them the time available prior to impact, available mission launch opportunities, the calculated probability of impact, and the time required to accomplish a successful deflection. Since the tracking data available for any given asteroid at issue can vary dramatically due to optical and radar tracking limitations, a deflection mission may well need to be deployed prior to a future impact being certain. In some instances (e.g. Apophis) a radio transponder will have to be sent to the asteroid in order to provide adequately accurate and timely information to rationally commit to a deflection[3]. In such instances the AT design can serve the dual role of first determining the precise orbit of the asteroid and therefore the need for deflection, and then, if a deflection is indicated, execute the mission. If, in this circumstance, a deflection is determined not to be required the AT spacecraft can, properly equipped, conduct an alternate asteroid characterization mission.

The technology readiness for an AT design depends strongly on the availability of high performance space electrical power. While solar electric propulsion (SEP) is currently available and proven, the more demanding NEO deflections will require greater power levels than can reasonably be provided by solar electric systems.

However, since for many NEOs at issue, a high performance deep space propulsion capability will be required in order to deliver any deflection system to the NEO, this propulsion system will also be available for use during the AT deflection. Such high performance deep space propulsion systems were recently under development in NASA's Prometheus Program but were canceled due to budget and priority problems.

The technical challenge specific to the AT (and any deflection concept envisioning surface operations) is the requirement for the spacecraft to attach to the asteroid surface after docking. While all forces associated with the deflection process are very light (perhaps 1 – 10 newtons) the current unknowns regarding the mechanical characteristics of NEO regolith make the design of an attachment system uncertain.

## II Mission Concept

The Asteroid Tugboat is conceptually very simple; rendezvous and dock with the asteroid a decade or more before the anticipated impact and then push it parallel to its velocity vector using the same propulsion system used to get to the asteroid until the deflection is completed. The simplicity of this concept is, however, complicated by the rotation of the asteroid which must be accounted for in order to thrust continuously parallel to the asteroid's velocity vector. To address this issue the AT must thrust at an angle to the vertical in order to develop a moment around the asteroid center of gravity thereby necessitating a secure spacecraft attachment to the asteroid's surface.

There are several conceptual alternatives which would obviate the need for such an attachment by requiring thrust only along the local vertical. One would be to dock in such a location that the local vertical would, due to the asteroid's rotation, approximately parallel the NEO velocity vector (and therefore the optimum thrust direction). Such an orientation, however, would for a short time occur only once per asteroid rotation, and in the general case, be limited to only a portion of the orbital arc at that. The exception to this latter would apply in the unique circumstance when the NEO spin axis was oriented perpendicular to the orbital plane. The very low thrusting duty cycle makes this alternative very unattractive.

A more complicated alternative would be to repetitively dock, thrust, lift off, reposition and redock, thereby keeping the thrusting episodes approximately aligned with the velocity vector. The operational complexity of this, in combination with limited duty cycle and the need to actively avoid boulders and other surface obstructions makes this alternative untenable.

Assuming then the availability of a secure anchoring capability the conceptually simplest solution to the rotation issue would be to dock on the NEO equator, thrust normal to the surface counter to the rotation until the spin rate drops to zero, then reorient and control the asteroid such that the thrust vector passes through the NEO center of mass and is parallel to the velocity vector and thrust continuously until the deflection is completed. Unfortunately the unintended consequence of zeroing the asteroid's rotation would be a likely gravitational restructuring of the asteroid (i.e. an asteroid-quake) seriously threatening the survival of the spacecraft. In fact any maneuver which alters the spin rate of a NEO is likely to cause a rearrangement of the asteroid since most of these bodies are thought to be "rubble piles," very loosely bound by weak gravitational forces balanced against weak rotational accelerations.

The resolution of this interesting challenge involves a sequence of maneuvers ultimately permitting the AT to thrust continuously at full capability parallel to the velocity vector.

Understanding the geometric sequence and phasing of this strategy is best achieved by visualizing the end point and working backward in time. The AT is docked at a rotational pole and attached. The final geometry for this method is (assuming the desire for a posigrade acceleration) such that the asteroid's angular momentum vector (i.e. spin axis) is located at a predetermined angle σ directly below the asteroid's velocity vector (i.e. the asteroid's velocity vector and the angular momentum vector define a plane perpendicular to the orbit plane). The AT then thrusts continuously parallel to the velocity vector, i.e. at an angle which is offset from the local vertical by the same angle as that between the spin axis and the orbital plane.

This off vertical thrust vector produces maximum continuous acceleration of the NEO in the desired direction while simultaneously generating a moment around the NEO center of gravity that torques the spin axis to maintain its position directly below the velocity vector as the asteroid proceeds around its orbit.

The geometry and detailed analysis of this method are described in detail in a technical paper written by Schweickart and Scheeres and presented at the 2004 AIAA Planetary Defense Conference[4]. A schematic diagram of the geometry is presented here as Figure 1.

This concept requires the AT engine to continuously rotate around the local vertical opposite to the asteroid's spin in order to maintain the thrust parallel to the velocity vector. This rotation is very slow however since most NEOs of interest have spin rates with periods greater than 2 hours.

**III Asteroid Tugboat Performance**

The capability of the AT is limited only by the electric propulsion system and the security of the asteroid attachment mechanism. The following example makes the assumption that, while certainly challenging, an engineering solution will be developed to provide secure attachment to asteroid surfaces once the mechanical characteristics of asteroid regolith are determined.

If one were to consider an ion or plasma propulsion system with a specific impulse ($I_{sp}$) of 10,000 seconds, a thrust of 2.5 newtons could be produced with an input power of 200-250 kW depending on the efficiency of the engine.

If one were to then consider the challenge of deflecting a 200 meter NEO using such a system it would require on the order of 4 months to achieve a change in velocity of 0.2 cm/sec, a value typical of the deflection velocity change required 10-20 years prior to impact.

Added to this time is that required for reorientation of the asteroid's spin vector. This challenge can vary dramatically with asteroid size due to $4^{th}$ power dependence of the moment of inertia on the asteroid's radius. For our 200 meter asteroid spinning 12 revolutions/day the time required for the reorientation maneuver would amount to less than one month if the spin axis had to be redirected by about 20 degrees.

While these numbers can vary considerably it is clear that the challenges we are most likely to face would easily fall into the realm of technical capability for a nuclear electric propulsion (NEP) system of the class considered for NASA's recently canceled Prometheus Program. If one assumes a nuclear-electric propulsion (NEP) system of the Prometheus class the AT would be capable of deflecting threatening NEOs on a direct Earth impact course (i.e. no intervening keyhole) up to 800 meters in diameter, if executed 10-20 years prior to impact.

### IV Mission Performance Considerations[5]

There are a host of additional considerations which come into play in considering a NEO deflection mission. It is critically important to keep in mind that such a mission falls into a completely different category from the normal scientific research or space exploration mission.

The following comments apply not only to the Asteroid Tugboat, but to any slowly-acting approach to deflection (serious but different considerations apply to instantaneous deflection or destruction approaches).

A NEO deflection will always be a public safety mission and will, without any doubt, be a major international event with unprecedented media attention. In many, if not most cases, the asteroid's positional error ellipse at the projected time of impact will still exceed the diameter of the Earth at the time when a deflection decision will have to be made, and hence the specific point of impact will not be known. The decision to deflect will often have to be made when the ultimate impact ground zero may be located in any of several countries spread across the face of the planet.

A NEO impact, even one for a specific NEO, is inherently an international affair and the demand for international coordination, if not authorization, will be strong. Given the possibility of failure during the course of deflection, there will be populations and property put at risk that will not have been at risk prior to the deflection operation. Such a potential failure implies a considerable financial liability on the part of the deflecting agency unless indemnified by pre-arrangement with the international community.

These, and other public concerns, argue strongly for there being a very high public confidence in the decision-making process, in the deflection methodology chosen, and in the agency executing the deflection.

With these and other considerations in mind it is critical that any deflection concept must, to be seriously contemplated,
1) be tested and demonstrated prior to use,
2) be capable of providing a precise and timely public announcement of the deflection result (i.e. resulting orbit),
3) be capable of providing assurance to the public that, if created, one or more large fragments do not continue to threaten an impact, and
4) be fully controllable in order that the NEO be targeted for a specific end state.

<u>Binary systems</u>: Another consideration, not yet well known to the public, is that a significant cohort of NEOs are binary (or multiple) systems. In many cases the secondary is itself large enough to penetrate the atmosphere and cause a threat to life and property. At this time it is thought that 15-20% of NEOs may be binary or multiple systems. In many cases the knowledge of whether or not this situation exists depends on obtaining a radar sighting of the NEO. Given that such radar sightings are rarely available and that the future funding for the powerful Arecibo radar is not assured, this challenge is doubly daunting.

Any deflection technique using an impulsive acceleration will not generally change the orbital track of a secondary. Furthermore unless there is pre-knowledge of a target NEO being a binary system, a separate and perhaps last minute additional mission may have to be mounted to deal with the situation.

Conversely the AT, due to the very low acceleration it imparts to the NEO, will simply cause the secondary to be dragged along with the primary during the deflection, *whether or not* it was known to exist prior to the mission.

<u>One mission vs. two</u>: Another consideration of considerable significance is the fact that an AT mission will fully rendezvous with the NEO at issue, and have aboard a radio transponder. These two facts produce substantial advantages for the AT over any impulsive deflection concept that does not complete a full rendezvous (i.e. match velocity with the NEO), and will provide public confidence in the conduct of the operation unavailable to most other deflection designs.

The criteria on which a NEO deflection decision will be based have not yet been developed. One key factor, however, will be the probability of Earth impact at the time when a mission must be launched to achieve a successful deflection. The probability of impact is directly related to the size of the "target" and inversely proportional to the size of the asteroid's uncertainty ellipse at the time of calculation. For a given deflection under consideration the "target" may be either the Earth itself or a resonant return keyhole associated with a close gravitational encounter preceding the impact. These two different targets can vary in size by many orders of magnitude. For example, in the case of Apophis the effective diameter of the Earth (accounting for gravitational focusing) is 27,600 km. while the width of the 2029 7/6 resonance keyhole is only 600 meters, over 45,000 times smaller! For 2004VD17 the ratio is not as extreme with the 2031 encounter keyhole being approximately 15 km wide and the Earth effective diameter 15,000 km.

In many cases available optical and radar tracking, as well as non-gravitational forces such as the Yarkovsky effect will result in a residual error ellipse, at the time when a deflection mission must be launched, that will result in considerable uncertainty whether the Earth will be hit at all. For example, in the Apophis case, which will be a very intensively tracked NEO by 2021 when a deflection mission would have to be launched (17 years of optical tracking and several radar apparitions) the size of the error ellipse will still be so large (~30 km) that even if the asteroid is headed for a direct impact with Earth the calculated probability of impact will be only 1 chance in 125, or less.

For this reason, Steve Chesley of JPL, who has done extensive analysis on this object[6] recommends that if there remains a non-zero impact probability following the 2013 observation of the NEO then a transponder should be deployed to Apophis to further reduce the error ellipse in support of a deflection decision for a 2021 launch.

If, at the time a deflection mission is being considered the residual uncertainty in the impact probability necessitates a transponder mission as a precursor, the use of an Asteroid Tugboat for this purpose is clearly desirable, if the requirements lie outside the capability of the Gravity Tractor. The requirements of the entire deflection sequence would be accomplished by a single spacecraft since on arrival at the asteroid the AT would first serve as the transponder tracking mission; then if, and only if, the NEO were determined to still be headed toward an impact it would shift into position and execute the needed deflection. For any impulsive deflection concept a separate transponder mission would have to be launched and then, if needed, a subsequent mission deployed for the deflection per se.

In a deflection maneuver any mission must plan to deflect the asteroid by a distance (at the time of impact) equal to at least the sum of half the best available residual error ellipse and half the "target" diameter. If the choice for the mission is an impulsive deflection (i.e. any mission which intercepts but does not match velocity with the NEO prior to its deflection operation) the required change in the NEO orbit will dramatically exceed the required change for an AT mission (or any concept using a full rendezvous) due to the dramatic reduction in the residual error ellipse as a result of the transponder tracking following arrival at the NEO.

Certainty in results: Finally, and in terms of public acceptance perhaps the most significant consideration of all, is the fact that the transponder aboard the AT spacecraft is available not only for the initial reduction of the remaining error ellipse, but throughout the deflection maneuver and more importantly, at the conclusion of the maneuver. There would be, as a result, no uncertainty as to whether or not the deflection was successful. It would have been tracked continuously from prior to the deflection, throughout the deflection maneuver itself, and at the completion of the deflection. In fact throughout the deflection the maneuver can be extended or otherwise modified in real time, based on ground tracking information. There will be no necessity to rely on assumptions about the response of the asteroid; full knowledge of the progress will be available in real time and adjustments can be made as necessary.

Other keyholes: One further subtle but important distinction remains between an AT/controlled deflection and an impulsive deflection, whether kinetic impact or explosive. That distinction resides in the existence of multiple resonance keyholes populating space nearby the Earth.

Unless a specific final orbit for the NEO to be deflected can be planned and the deflection itself can be shown to have achieved this plan, the public cannot be assured that the NEO itself, or large fragments of it, will not have ended up headed for another keyhole thereby still threatening the Earth[8]. Claims that 'we think it went successfully' will not be adequate.

For example, until fairly recently the error ellipse for Apophis, despite two years of tracking, contained several resonant return possibilities, including the 8/7 and 15/13 keyholes. Unless a deflection is controlled, i.e. unless it can both target a desired endpoint and guide to it, the result of an impact may well be to simply shift the impact a year or two or ten. Furthermore, unless there is an immediate precise determination of the post deflection orbit it may well take considerable time and tracking before the general public can be assured that in fact the asteroid, or a major fragment (in the case where impact or explosion is used) is not still headed for Earth impact.

## V Conclusions

An Asteroid Tugboat deflection is fully controlled providing an accurate final determination of the need for a deflection, the ability to target for and achieve a specific safe final orbit, and precise and immediate knowledge of the final result. No alternative impulsive deflection concept can provide these capabilities.[7]

The Asteroid Tugboat ultimately depends on the availability of a high performance deep space propulsion system. However, for NEOs which require a launch vehicle capability beyond that currently available in the inventory (i.e. those in eccentric and/or highly inclined orbits) a development program for high efficiency propulsion will be required in any event. This requirement exists regardless of deflection concept since no system can begin its work until first being delivered to the NEO[8]. Where advanced electric propulsion (NEP or other) is required for the intercept with the NEO the AT will utilize the power source and engines of the spacecraft for the deflection operations as well.

The primary current limitation to the use of the Asteroid Tugboat is the lack of knowledge necessary to design and engineer an attachment system to securely hold the spacecraft to the asteroid surface. While the maximum forces generated by the AT throughout its operations are very light (~ 0.5 lbs.) these forces must be applied at angles up to 90 degrees from the local vertical in order to generate rotational moments to the asteroid. The information required for designing the attachment mechanism can only be obtained from in-situ measurements on the surface of one or more asteroids. It is therefore critical that one or more asteroids in the size range of interest be visited and the surface regolith characterized.

Finally it is very critical that neither NASA nor any other agency involved in addressing this challenge underestimate the degree to which the international community, both at the state level and that of the general public, will demand to be involved in and ultimately satisfied with many of the decisions regarding NEO deflection. Fragmentation of the NEO, uncertainty in the execution and the results, and even nuclear explosions and radiation will be of enormous concern. Where more certain and benign methods are available to accomplish the deflection such instantaneous but risky approaches will not be acceptable.

The Asteroid Tugboat, when capable of meeting the challenge, will be both technologically and societally preferable to impulsive techniques which produce uncertain results.

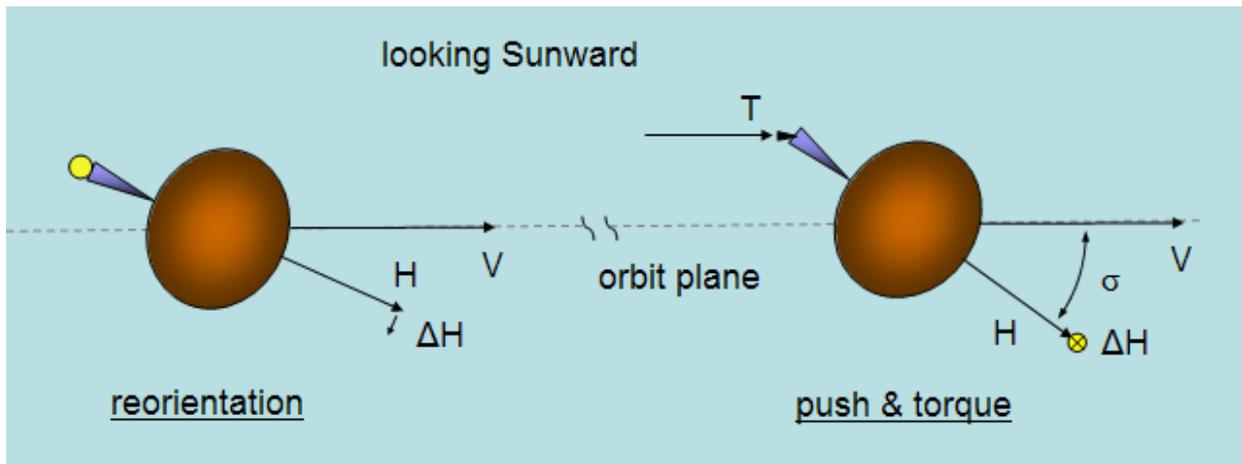

**Figure 1. A schematic representation of the Asteroid Tugboat operation for deflecting an asteroid. After docking with and attaching to the asteroid surface on the spin axis the reorientation phase is begun by thrusting horizontally (into or out of the page) to create a moment which increases or decreases the angle (σ) of the NEO spin vector with respect to the orbit plane to a predetermined value. Subsequently with the NEO angular moment vector (H) directly below the velocity vector (V) the Asteroid Tugboat thrusts parallel to V until the desired change in velocity is achieved. The depression angle (σ) is calculated such that the moment caused by the thrusting maneuver continues to torque the H vector thereby maintaining it directly below the velocity vector throughout the deflection.**

---

[1] Russell L Schweickart, B612 Foundation; Piet Hut, Institute for Advanced Study; Clark Chapman, Dan Durda, Southwest Research Institute.
[2] *The Asteroid Tugboat*, Schweickart, Lu, Hut, and Chapman, Scientific American, November 2003, (See #3, http://www.b612foundation.org/press/press.html)
[3] *Potential Impact Detection for Near-Earth Asteroids: The Case of 99942 Apophis (2004 MN4)*, Steve Chesley, Asteroids, Comets, Meteors Proceedings, IAU Symposium No. 229, 2005 (See #11, http://www.b612foundation.org/press/press.html)
[4] *The Mechanics of Moving Asteroids*, AIAA Conference on Planetary Defense, Feb 2004 (see http://www.b612foundation.org/press/press.html, #4)
[5] This section of the white paper is essentially identical with the comparable section for the Gravity Tractor deflection paper since both designs address these mission performance characteristics.
[6] *Potential Impact Detection for*

*Near-Earth Asteroids: The Case of 99942 Apophis (2004 MN4)*, Steve Chesley, Asteroids, Comets, Meteors Proceedings, IAU Symposium No. 229, 2005 (See #11, http://www.b612foundation.org/press/press.html)

[7] There are several conceptual non-impulsive concepts which share these capabilities, but they appear to be far more costly and/or dauntingly complex.

[8] The exceptions to this are those dramatic but risky techniques where a rendezvous is not conducted. Rather a direct impact or a nuclear explosive is set off at precisely the right time as it passes by the NEO. Neither of these schemes can assure the public that there are not large fragments still headed for impact, nor in fact what precisely was the outcome of the operation.